\documentclass[prd,twoside,twocolumn,superscriptaddress,nofootinbib]{revtex4-1}
\usepackage{amsmath,amssymb}
\usepackage{graphicx}
\usepackage{xcolor}
\usepackage{multirow}
\usepackage{cancel}
\usepackage{color}
\usepackage{enumitem}
\usepackage{slashed}

\usepackage[colorlinks,citecolor=blue]{hyperref}

\newcommand{\be}{\begin{equation}}
\newcommand{\ee}{\end{equation}}
\newcommand{\beq}{\begin{equation}}
\newcommand{\eeq}{\end{equation}}
\newcommand{\bea}{\begin{eqnarray}}
\newcommand{\eea}{\end{eqnarray}}
\newcommand{\besp}{\begin{equation}\begin{split}}
\newcommand{\eesp}{\end{split}\end{equation}}

\newcommand{\Eq}[1]{Eq.~(\ref{#1})}

\newcommand{\Dfbd}{\mathord{\buildrel{\lower3pt\hbox{$\scriptscriptstyle\leftrightarrow$}}\over {D}_{\mu}}}

\def\mG{\mathcal{G}}
\def\mH{\mathcal{H}}

\def\mO{\mathcal{O}}

\def\0{\textbf{0}}
\def\1{\textbf{1}}
\def\2{\textbf{2}}
\def\3{\textbf{3}}
\def\4{\textbf{4}}
\def\5{\textbf{5}}
\def\6{\textbf{6}}
\def\7{\textbf{7}}
\def\8{\textbf{8}}
\def\9{\textbf{9}}

\def\q{\textbf{q}}

\begin{document}

\title{Broad composite resonances and their signals at the LHC }

\author{Da Liu}
\email{da.liu@anl.gov}
\affiliation{High Energy Physics Division, Argonne National Laboratory, Argonne, IL 60439, USA}

\author{Lian-Tao Wang}
\email{liantaow@uchicago.edu}
\affiliation{Department of Physics and Enrico Fermi Institute, The University of Chicago, Chicago, IL 60637, USA}

\author{Ke-Pan Xie}
\email{kpxie@snu.ac.kr}
\affiliation{Center for Theoretical Physics, Department of Physics and Astronomy, Seoul National University, Seoul 08826, Korea}

\begin{abstract}
The existence of the $SU(2)_L$ triplet composite spin-1 resonances $\rho^{\pm,0}$ is a universal prediction of the strongly interacting new physics  addressing the naturalness problem. Such resonances have not been found  in the di-boson final states,
which are expected to be the dominant decay channels. In this work we propose a new scenario where the left-handed quark doublet $q_L = (t_L, b_L)$ is fully composite.  In this case, the $\rho$-resonances can be broad and mainly decay to the third generation quarks. The $t\bar{t}$ resonance search channel is comparable in sensitivity to the di-lepton channel. In addition,  the same-sign di-lepton channel in the $t\bar{t}\rho^0$ associate production  can probe the large width region and complementary to the Drell-Yan production channels.
\end{abstract}

\maketitle

{\bf Introduction.}--The discovery of a Higgs-like boson at the LHC~\cite{Aad:2012tfa,Chatrchyan:2012xdj} was a big step towards the understanding of the electroweak symmetry breaking (EWSB). An attractive solution to the associated naturalness problem is provided by the composite Higgs models, in which the Higgs is a pseudo Nambu-Goldstone boson emerged from the spontaneous symmetry breaking $\mG/\mH$ of some strong interacting composite sector at $\mO({\rm TeV})$. The EWSB is triggered by some explicit $\mG$-breaking interactions between elementary SM sector and composite sector~\cite{Kaplan:1991dc,Contino:2003ve,Agashe:2004rs}. An important signal of composite Higgs models is the presence of composite resonances. A spin-1 resonance similar to the $\rho$ of QCD (denoted also as $\rho$ here) is probably one of the most obvious targets for collider searches. In most previous studies, the $\rho$-resonances tend to be narrow and decay dominantly to the SM di-boson final states (i.e. $W^\pm Z/W^\pm h$, $W^+W^-/Zh$)~\cite{Thamm:2015csa,Pappadopulo:2014qza}.
In this article, we propose a new scenario where the left-handed third generation quark doublet $q_L=(t_L,b_L)^T$ is a massless bound state from the composite sector. The $\rho$-resonances can be broad and mainly decay to the third generation final states (i.e. $t\bar t$, $b\bar b$, $t\bar b/\bar tb$). The search for spin-1 composite resonances will be more challenging, as will be discussed in this work. Currently, the LHC experimental collaborations are planning for the new search strategies for the future, including the high luminosity upgrade. Our paper emphasizes a new direction,  the search for broad resonances. We present a benchmark model in the framework of composite Higgs models, offering timely motivation and important targets for testing the strategies in this new direction of the LHC searches.

The symmetry structure of a composite Higgs model is a coset $\mG/\mH$, where the strong dynamics preserves $\mH$ even after confinement. Hence, the composite resonances should fill complete multiplets of $\mH$. The Standard Model (SM) gauge groups are embedded in $\mH$. 
In popular benchmarks of the composite Higgs model, the SM fermions are treated as elementary. The third generation fermions acquire their masses through mixing with composite fermions~\cite{Kaplan:1991dc,Contino:2003ve,Agashe:2004rs}. In this scenario, the coupling between the SM fermions and the $\rho$ is suppressed by either $\rho$-SM gauge boson mixing or 
elementary-composite fermion mixing. The $\rho$-resonances couple strongly to other composite states, such as 
the longitudinal modes of the $W$ and $Z$,  and the Higgs boson. However, the lack of color factor enhancement and some accidental small factor lead to the decay width-mass ratio of the $\rho$  to be $\mO(g_\rho^2/96\pi)$. Hence, the $\rho$-resonance appears to be narrow even if $g_\rho$ is sizable.

Heaviness of the third generation fermions motivates considering some of them as fully composite, as part of a complete multiplet of $\mH$. Previously,  $t_R$ has often been treated as fully composite singlet of $\mH$~\cite{DeSimone:2012fs,Marzocca:2012zn}. Though simple, this is not the only possibility. We will consider more extended possibilities. This gives very different predictions for the width of the $\rho$-resonances, which lead to qualitatively new features and challenges for collider searches.

{\bf The Model.}--We present here a model which realizes the new features discussed in the Introduction. Beginning with the frequently used coset $SO(5)/SO(4)$, we consider the left-handed third generation quarks as fully composite,  embedded as a $\4$ of $SO(4)$,
\bea
\Psi_L& =& \frac{1}{\sqrt{2}}\left(
\begin{array}{c}
i b_L - i X_L \\
 b_L + X_L \\
i t_L + i T_L \\
 -t_L + T_L 
\end{array}
\right)_{2/3} =P\begin{pmatrix}q^X_L\\q_L\end{pmatrix}, 
\eea
where $P$ is a $4\times 4$ unitary matrix. Under the decomposition $SO(4)\times U(1)_X\to SU(2)_L\times U(1)_Y$, the quartet can be decomposed as bi-doublets,
\be
\4_{2/3}\to \2_{7/6}\oplus\2_{1/6}.
\ee
$q_L=(t_L,b_L)^T$ and $q_L^X=(X_L,T_L)^T$ have the SM quantum number $(\3,\2)_{1/6}$ and $(\3,\2)_{7/6}$, respectively. We assume that the right-handed top quark $t_R$ is elementary. We also introduce an elementary doublet $q_R^X=(X_R,T_R)^T$ with SM quantum number $(\3,\2)_{7/6}$, which pairs up with $q_L^X$ and becomes massive~\cite{Agashe:2005vg}. We will write them as incomplete $\5$ of $SO(5)$, $t_R^\5=(0 , 0 ,0, 0, t_R)_{2/3}^T$ and $q_R^{X\5}=1/\sqrt2(-iX_R, X_R, iT_R, T_R,  0)_{2/3}^T$. The spin-1 composite resonances, the $\rho$s, span the adjoint of $SO(4) = SU(2)_L \times SU(2)_R$. We mainly focus on $\3$ of $SU(2)_L$, $\rho^{a_L}$, with $a=1,2,3$. The relevant Lagrangian, following the standard Callan-Coleman-Wess-Zumino procedure~\cite{Contino:2011np,Greco:2014aza,Liu:2018hum}, is
\bea
\label{eq:LagrhoL}
\mathcal{L}=&-&\frac{1}{4}\rho_{\mu\nu}^{a_L}\rho^{a_L\mu\nu}+\frac{m_{\rho}^2}{2g_{\rho}^2}(g_{\rho}\rho_\mu^{a_L}-e_\mu^{a_L})^2  +\bar q_{R}^Xi\slashed{D}q_{R}^X\nonumber\\
&+&\bar t_{R}i\slashed{D}t_{R}+\bar \Psi_L\gamma^\mu\left(i\nabla_\mu+\frac{2}{3}g_1B_\mu\right)\Psi_L  \\
&+&c_1\bar \Psi_L\gamma^\mu T^{a_L}\Psi_L(g_{\rho}\rho^{a_L}_\mu-e^{a_L}_\mu) \nonumber\\
&-&y_{1R}f \bar q_R^{X\5}U\Psi_L-y_{2R}f \bar t_R^{\5}U\Psi_L+\text{h.c.} + \sum_i\alpha_iQ_i, \nonumber
\eea
where $\nabla_\mu=\partial_\mu-ie_\mu^{a_L}T^{a_L}-ie_\mu^{a_R}T^{a_R}$. The field strength tensor is $\rho_{\mu\nu}^{a_L}=\partial_\mu\rho_\nu^{a_L}-\partial_\nu\rho_\mu^{a_L}+g_\rho\epsilon^{abc}\rho_\mu^{b_L}\rho_\nu^{c_L}$. $U=\exp\left\{i\frac{\sqrt{2}}{f}h_iT^{\hat i}\right\}$ is the Goldstone matrix, with $T^{\hat i}$ being the generators of $SO(5)/SO(4)$. The $Q_i$s are a set of higher-order operators~\cite{Contino:2011np}. The top quark mass $M_t$, the top partner masses $M_{T,X}$,  are given by:
\beq
M_t \sim \frac{y_{2R}v}{\sqrt{2}}, \qquad M_X = y_{1R} f, \qquad M_T \sim   y_{1R} f.
\eeq
There are various indirect constraints on this model.  

{\bf The anomalous couplings.}--The first set of constraints come from the modification of the  $Z b_L\bar{b}_L$, $Z t_L \bar{t}_L$, $W t_L \bar{b}_L$ couplings. The composite fermion kinetic term has an accidental  $P_{LR}$ parity symmetry, which exchanges $T^{a_L} \leftrightarrow  T^{a_R}$, $e_\mu^{a_L} \leftrightarrow e_\mu^{a_R} $~\cite{Agashe:2006at,Contino:2011np}. This protects the $Z b_L \bar b_L$ coupling at tree level, avoiding dangerous deviations with the sizes of $\xi \equiv v^2/f^2$. 
More explicitly, the Lagrangian in \Eq{eq:LagrhoL} contains two potentially dangerous operators 
$\mO_L^q=\bar q_L\gamma^\mu q_L(H^\dagger i\Dfbd H)$, and $ \mO_L^{(3)q}=\bar q_L\gamma^\mu\sigma^a q_L(H^\dagger\sigma^a i\Dfbd H)$. They can modify the $Zb_L \bar b_L$ coupling as $\delta g_{Lb}=-\frac{v^2}{2}(c_L^{q}+c_L^{(3)q})$~\cite{Gori:2015nqa}. In our model, these contributions cancel since $c_L^{q}=-c_L^{(3)q}=1/(4f^2)$, as shown in \Eq{eq:LagrhoL}. However, the mass terms in \Eq{eq:LagrhoL} do not preserve the $P_{LR}$, which corrects the $Zb_L \bar b_L$ coupling  at one loop level. We have checked that their constraints are weaker than the $S$, $T$-parameters~\cite{Panico:2010is,Delaunay:2010dw}. The modifications to the $Wt_L\bar b_L$, $Z t_L \bar{t}_L$ couplings arise at tree level, given by $\delta g_{Wt_Lb_L} \sim \delta g_{Zt_Lt_L} \sim -\xi/4$~\cite{Efrati:2015eaa}. The bond from the electroweak precision test (EWPT) is $|\delta g_{Zt_Lt_L}| \lesssim 8\%$~\cite{Panico:2018fug,Efrati:2015eaa}, which limits $\xi\lesssim 0.32$. The newest limit of the $Zt_L\bar t_L$ coupling  from $t\bar tZ$ associated production at the 36.1 fb$^{-1}$ LHC is  $|\delta g_{Zt_Lt_L}| \lesssim 10\%$ (95\% C.L.)~\cite{ATLAS-CONF-2018-047}, corresponding to $\xi\lesssim0.4$.

{\bf Oblique parameters.}--In our model, the strong dynamics preserves a $SO(4)$ symmetry. Since $SO(4)$ contains the custodial $SU(2)$, there is no tree level contribution to the $T$-parameter. The $S$-parameter receives a tree level contribution from the mixing of $\rho$ and SM gauge bosons. One loop contributions to the $T$ and $S $ come from heavy quarks, $\rho$-resonances, and the modified Higgs-gauge boson couplings. Since our model in \Eq{eq:LagrhoL} is non-renormalizable, the loop contributions to $S$, $T$ are in principle incalculable. We regulate the divergence by a cutoff $\Lambda = 4 \pi f$.  We have used the results of~\cite{Ghosh:2015wiz,Contino:2015mha,Contino:2015gdp} the contribution from the loop of $\rho$s, and calculated the contributions from fermion loops using the formulae in~\cite{Lavoura:1992np}.

\begin{figure}
\centering
\includegraphics[scale=0.3]{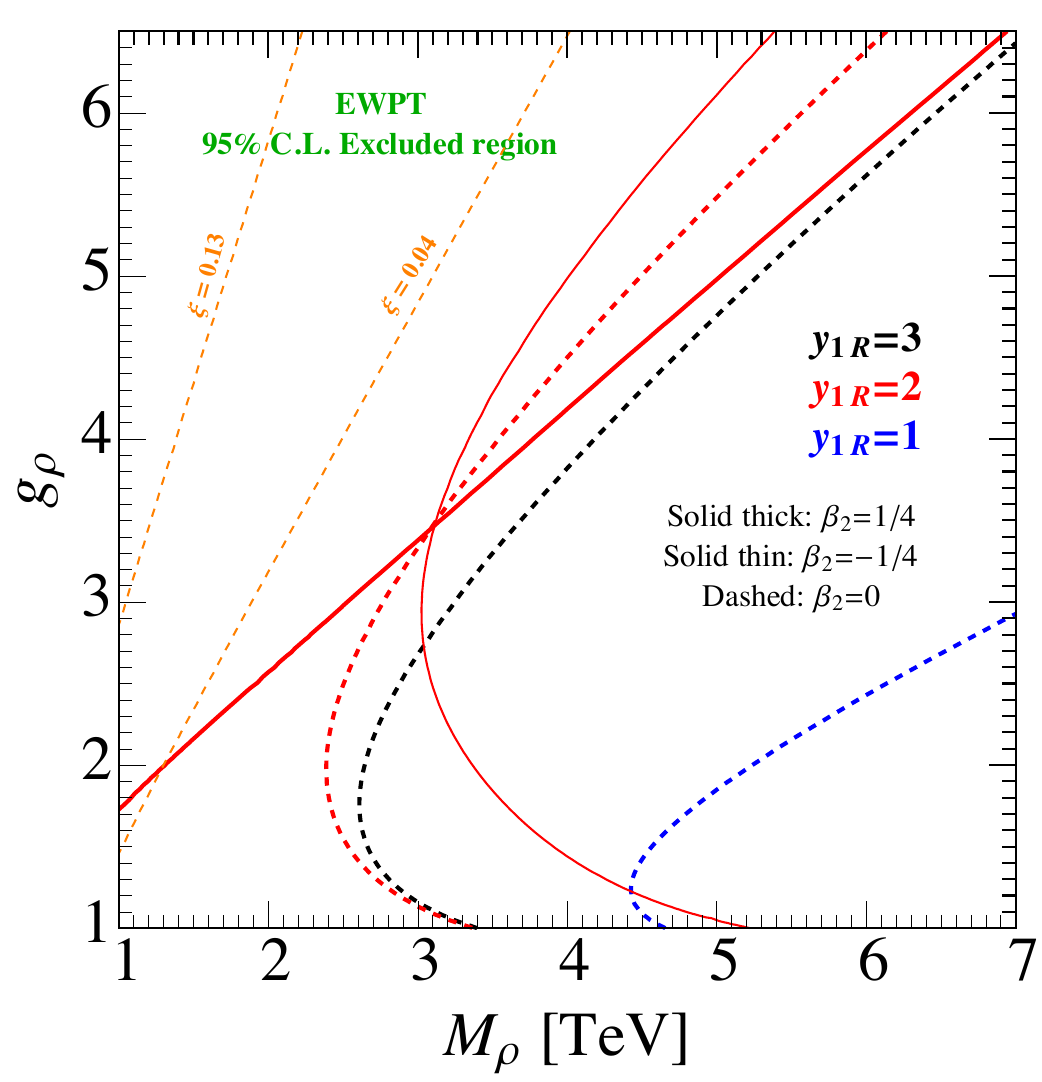}
\caption{The indirect bound (95\% C.L.) from EWPT in the $M_\rho -g_\rho$ plane for $a_\rho = 1/2$ with different parameters of $y_{1R}$ and $\beta_2$. The orange contours $\xi=0.13$ and 0.04 represent the indirect bounds at the current and 3 ab$^{-1}$ LHC~\cite{Sanz:2017tco,deBlas:2018tjm,CMS-NOTE-2012-006,ATL-PHYS-PUB-2013-014,Dawson:2013bba}, respectively.}
\label{fig:EWPT}
\end{figure}

There can also be additional contribution to the $S$-parameter from higher order operators. An operator which is particularly relevant is
\be
Q_2=g_\rho\rho^{a_L}_{\mu\nu}E^{a_L\mu\nu},
\ee
where $E_{\mu\nu}^{a_L}$ is given by~\cite{Panico:2015jxa}, and 
\bea
E_{\mu\nu}^{a_L}&=&\cos^2\frac{|\vec{h}|}{2f}g_2W^{a_L}_{\mu\nu} \nonumber\\
& &-\frac{4}{|\vec{h}|^2}\sin^2\frac{|\vec{h}|}{2f}\vec{h}^Tt^{a_L}(g_1B_{\mu\nu}t^{3_R})\vec{h},
\eea
in the unitary gauge. It can contribute to the mixing between $\rho$ and SM  gauge bosons, and hence shift the $S$-parameter. 
According to the so-called partial UV completion assumption, $\alpha_2\lesssim1/g_\rho^2$~\cite{Contino:2011np}. Therefore, the kinetic mixing between $\rho_\mu$ and $W_\mu$ induced by $Q_2$ is sub-leading. We often define $\beta_2=g_\rho^2\alpha_2$, with $\beta_2$ being a $\mO(1)$ parameter.
There is no similar contribution to the $T$-parameter, if we assume custodial symmetry is preserved in the UV completion. In Fig.~\ref{fig:EWPT},  we have plotted the 95\% C.L. bound from $S$, $T$ measurement on the $M_\rho - g_{\rho}$ plane with $a_\rho = m_\rho/(g_\rho f) =1/2$ with different values of $\beta_2$, $y_{1R}$ (see Appendix~\ref{app} for the analytical formulae for the $S$, $T$ in our model at leading order in $\xi$), using the limits on $S$ and $T$ in Ref.~\cite{Tanabashi:2018oca}:
\beq
 S = 0.02 \pm 0.07, \qquad T = 0.06 \pm 0.06, 
\eeq
with strong correlation 92\%. Note that  the mass of the top partner  is roughly given by $y_{1R}f$. Small $y_{1R}$ will lead to strong constraint on $f$ due to  contributions to the $S$, $T$ of $\mO(M_t^2/M_T^2)$. Meanwhile, $y_{1R}$ explicitly breaks the custodial symmetry and larger $y_{1R}$ will lead to stronger constraint.  We find that $y_{1R} \sim 2$ gives the  weakest bound on our parameter space, as can be seen from the figure. $\beta_2 = 1/4$ significantly relaxes bounds in the small $M_\rho$ region, as the tree level contributions to the $S$-parameter from the higher dimension operator $Q_2$ and the mass term cancel. The bounds come from EWPT can be further relaxed if there is new positive contribution to the $T$-parameter~ \cite{Ghosh:2015wiz,Contino:2015mha,deBlas:2016nqo,Ciuchini:2014dea,Ciuchini:2013pca}. 

Finally, we briefly comment on the flavor physics implications. The main issue is the mass difference of the $B$ and $\bar B$ mesons, which requires~\cite{Giudice:2007fh}
\be
\xi c_{4q}\left(\frac{\theta_{bd}}{V_{ub}}\right)^2<2\times10^{-3},
\ee
where $\theta_{bd}$ is the projection of $b_L$ into the $d$ mass eigenstate, and $c_{4q}$ is the Wilson coefficient of the operator $(1/f^2)\bar q_L\gamma^\mu q_L\bar q_L\gamma_\mu q_L$. In general, this strongly constrains the fully composite left-handed top quark scenario~\cite{Gillioz:2008hs,Barbieri:2007bh}. However, this constraint is very dependent on the underlying theory of flavor. For example, the CKM matrix can be originated from the up-type quark sector. In this case,  the down-type sector is flavor diagonal, i.e. $\theta_{bd}=0$, making the model free of the $B$-physics constraint.

\begin{figure}
\includegraphics[scale=0.4]{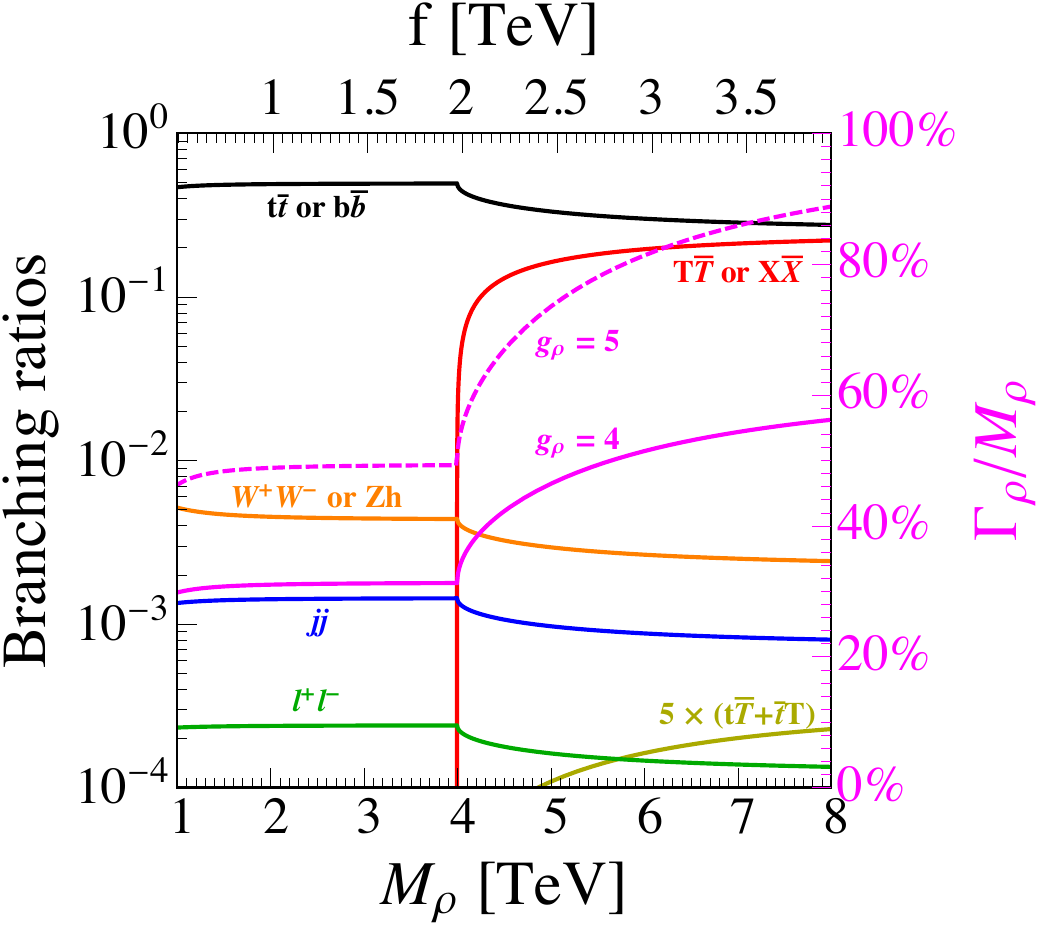}
\caption{The decay branching ratios of $\rho$ and the ratio of its total width in comparison and mass. We set $g_\rho=4$, $a_\rho=1/2$, $M_X=2$ TeV and $c_1=1$  for the branching ratios. For $\Gamma_\rho/M_\rho$, we plot two magenta curves with $g_\rho=4$ (solid) and 5 (dashed). }
\label{fig:width}
\end{figure}

{\bf Collider signal.}--The most significant difference between the collider signal of the spin-1 composite resonances in our model and those of the previously used benchmarks is the width. The branching ratios into different final states  and the total decay width for the neutral resonance $\rho^0$ are shown in Fig.~\ref{fig:width}. Since $q_L$ is fully composite, its coupling to the $\rho$ is of the order $g_\rho$. The dominant decay channels are $t\bar{t}$, $b\bar{b}$  for $M_\rho < 2 M_X$.  If $M_\rho > 2 M_X$, the decay into top partner pair is significant, which is almost half of the total decay widths in this region.  Broad $\rho$-resonances caused by the decay to top partners were studied in Refs.~\cite{Barducci:2012kk,Barducci:2015vyf}. The branching ratio  to the di-boson final state is suppressed by a factor of $a_\rho^4/(2N_c)$.  The suppression of the di-boson branching ratio, especially at small $a_\rho$, makes them much less relevant. This is very different from the well-studied cases, where the di-boson channel is the most sensitive~\cite{Pappadopulo:2014qza}.

\begin{figure}
\centering
\includegraphics[scale=0.35]{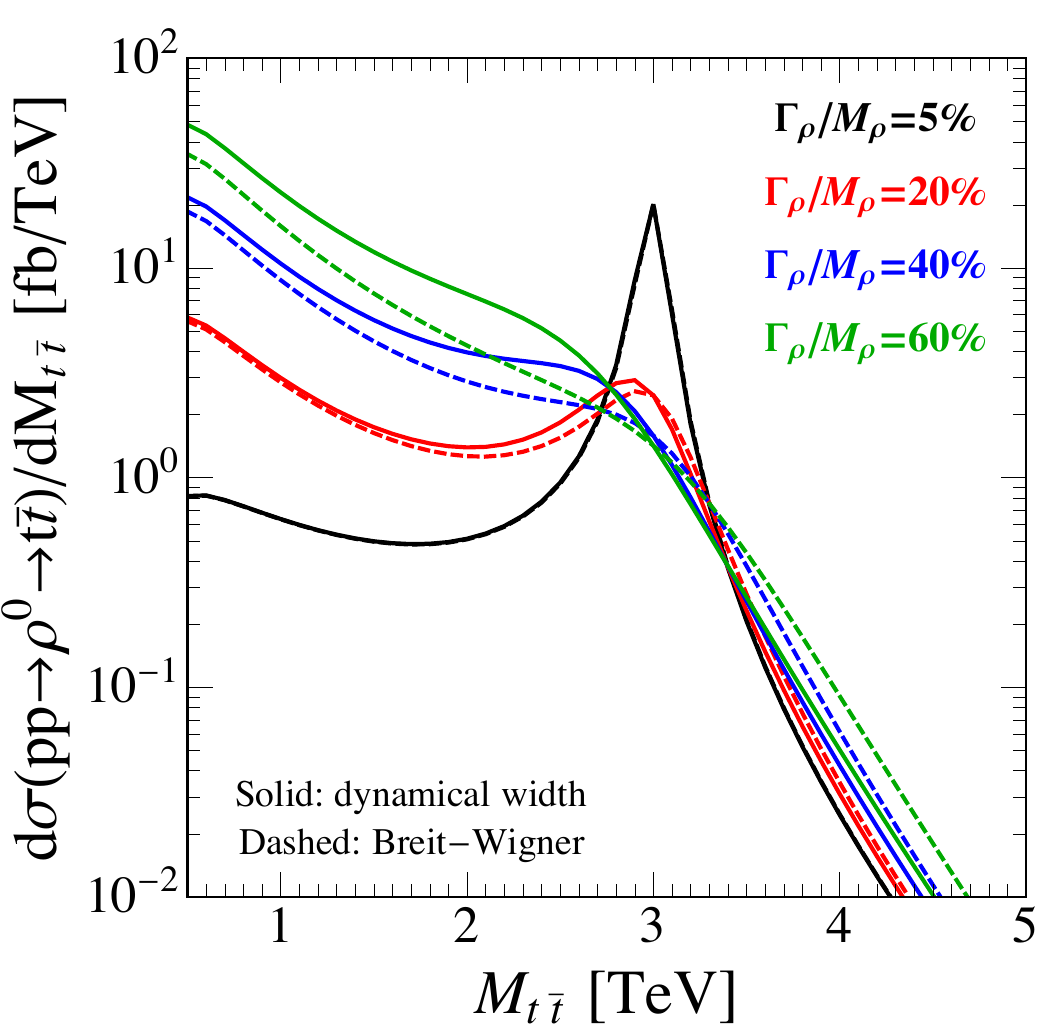}
\caption{Shape of the broad composite resonance at the LHC. 
We set $M_\rho=3$ TeV, $a_\rho=1/2$, $y_{1R}=2$ and $c_1=1$. }\label{Invariant_mass}
\end{figure}

For broad resonances, the usual narrow width approximation does not apply. Nor is it correct to just add a large constant width to the propagator.  Instead, we need to replace the propagator
\be
\label{eq:running}
\frac{1}{(\hat s-M_\rho^2)^2+M_\rho^2\Gamma_\rho^2}\to \frac{1}{(\hat s-M_\rho^2)^2+\hat s^2\Gamma_\rho^2/M_\rho^2},
\ee
where $\sqrt{\hat s}$ is the parton center of mass energy. This has a significant impact on the shape of the resonance at the LHC, as shown in Fig.~\ref{Invariant_mass}. 

\begin{figure}
\includegraphics[scale=0.45]{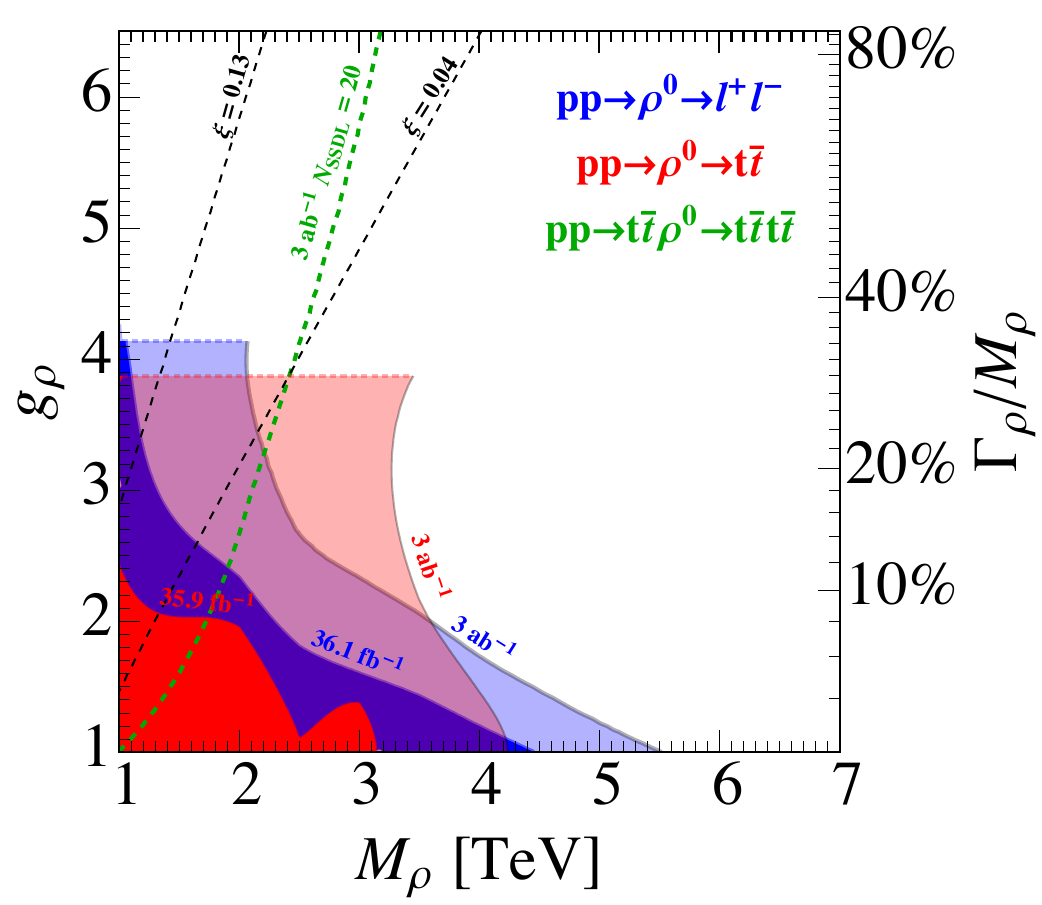}
\caption{The current and projected constraints. The parameter benchmarks are $y_{1R}=2$, $a_\rho=1/2$ and $c_1=1$. The $t\bar t$ and $\ell^+\ell^-$ bounds are based on the CMS~\cite{Sirunyan:2018ryr} and ATLAS~\cite{Aaboud:2017buh}, respectively. For the $t\bar t\rho^0\to t\bar tt\bar t$ channel, we use the SSDL event contour $N(\ell^\pm\ell^\pm+{\rm jets})=20$ to set an estimate for the 3 ab$^{-1}$ LHC.}\label{fig:exclusion}
\end{figure}

There is no LHC search fully optimized for the broad resonances presented here. Achieving maximal sensitivity will be a challenge which deserves much more detailed studies.  In the following, we will recast some of the LHC searches which still have sensitivity and highlight the difference with the well studied benchmarks~\footnote{Another dedicated recast for a similar scenario can be found in a recent study~\cite{Gintner:2019fvr}.}. 
First of all,  the searches in di-boson channel are not sensitive due to its suppressed branching ratio.  Moreover, the limit set by searching for narrow resonances in the $t \bar t$, $b\bar{b}$, $t\bar b/\bar tb$ and $\ell^+\ell^-$ final states will not apply if $\Gamma_\rho/M_\rho > 40\%$. The systematic uncertainties on the backgrounds will have a large impact for the large width case. There are several broad resonance searches at the LHC in the above channels, but most of the searches have used the constant decay width approximation which could mis-model the signal. For the $t\bar t$ channel, the large width effect has been considered up to $\Gamma_\rho/M_\rho\sim 30\%$ by the ATLAS~\cite{Aaboud:2018mjh} and the CMS~\cite{Sirunyan:2018ryr}. While ATLAS searched in the semi-leptonic final state, the CMS analysis combines all possible final states and is more sensitive. In Fig.~\ref{fig:exclusion},  we show the present limits and projected (3 ab$^{-1}$) reach for the $t\bar t$ channel (red shaded region) based on the CMS result. The colored regions are truncated at $g_\rho\sim4$ ($\Gamma_\rho/M_\rho\leqslant 30\%$), beyond which reliable extrapolations from current searches are not possible. When $g_\rho$ increases, the reach of $M_\rho$ first decreases because of  the suppression of the coupling between the $\rho$ resonance and valence quarks at large $g_\rho$. It then increases as the $b\bar{b}$ initiated production becomes important. The possibility of a broad $\rho^0$ decaying into $\ell^+\ell^-$ has been studied by ATLAS~\cite{Aaboud:2017buh}, up to $\Gamma_\rho/M_\rho=32\%$. The corresponding limit and its extrapolation to 3 ab$^{-1}$  are shown in Fig.~\ref{fig:exclusion} (blue regions). The mass reach in low $g_\rho$ region is higher than the $t\bar t$ channel. While the high $g_\rho$ region, due to the branching ratio suppression, $\ell^+\ell^-$ is worse. 

Currently, there is no strong constraint from the  $b\bar b$ channel. ATLAS has searched for a broad $b\bar b$ resonance up to $\Gamma_\rho/M_\rho=15\%$~\cite{Aaboud:2016nbq}, but the constraint is too week to be shown in Fig.~\ref{fig:exclusion} due to the low integrated luminosity (3.2 fb$^{-1}$). CMS searched in the di-jet channel for  both the narrow and broad resonances~\cite{Sirunyan:2018xlo}. The study considers the dynamical width effect, and gives results for $\Gamma_\rho/M_\rho$ up to 30\%. Without $b$-tagging in this search, its limit is weak.

Besides the Drell-Yan processes, there are other sensitive channels. Since the left-handed top is strongly coupled, the same-sign di-lepton (SSDL) channel in the four top final state $pp\to t\bar{t}\rho^0\to t\bar{t}t\bar t$ can be useful~\cite{Liu:2015hxi}. This channel has mild dependence on the modeling of the width. The estimated sensitivity in our parameter space set by requiring 20 SSDL signal events (the green contour) is shown in Fig.~\ref{fig:exclusion}. This channel can cover the large $g_\rho$ region, which is hard to probe via Drell-Yan process. 

The signature in the large coupling region $g_\rho \gtrsim 4$ would be very broad heavy resonance in the $\ell^+\ell^-$, $t\bar{t}$, $b\bar{b}$, $t\bar{b}/\bar{t}b$ final states. One possible way to enhance the sensitivity of is to consider the interference between the signal and the SM irreducible background. This is similar to explore energy growing behavior from the higher dimension (four fermion) operators~\cite{Franceschini:2017xkh,Liu:2018pkg,Alioli:2017nzr,Alioli:2017jdo,Farina:2016rws,Domenech:2012ai,Pomarol:2008bh,Bellazzini:2017bkb,Kelley:2010ap,Banerjee:2018bio}. Since our $\rho$ resonance is color neutral, the Drell-Yan production channel does not interfere with the QCD $t\bar{t}$ background. The $t$-channel $bb\to bb$ and the Drell-Yan $\ell^+\ell^-$ do have interference with the SM irreducible backgrounds. Due to the suppression of bottom PDF at high energy and the suppression of the di-lepton branching ratio at large $g_\rho$, they don't have significant sensitivity to the region $M_\rho \gtrsim 4$ TeV, $g_\rho \gtrsim 4$. The productions of top partners $T$ and $X$ can probe our model. Compare to the pair production, the single production of $T$, $X$ can reach a higher mass region. We estimate the sensitivity in the SSDL channel from singly produced of the charge-5/3 top partner $X$, requiring $N(\ell^\pm\ell^\pm+{\rm jets})=20$. It can reach $M_X\sim2.6$ TeV (corresponding to $\xi\sim0.036$, with $y_{1R} = 2$, $a_\rho = 1/2$, $c_1 = 1$). 
Additional handles on the signal could become important while reconstruction of a sharp resonance is less effective. 
For example, the $\rho$ resonance strongly interacts with the left-handed top quarks, the polarization measurement of the top quarks in the $t\bar{t}$ final states can also help improve the sensitivity. It is well known that in the top quark rest frame, the polar angle $\theta_\ell^*$ distribution of the charged lepton from the decay $t\to b\ell^+\nu$ reflects right- (left-) handed polarization of the top quark, i.e. $dN/d\cos\theta_{\ell}^*\sim 1\pm\cos\theta_{\ell}^*$~\cite{Grzadkowski:2001tq,Grzadkowski:2002gt,Godbole:2006tq,Godbole:2009dp,Godbole:2010kr,Godbole:2018wfy}. Since the QCD-produced top pairs are unpolarized, this distribution asymmetry can be used to distinguish our signal from the background. See, for example Refs.~\cite{Agashe:2006hk,Cerrito:2016qig} for the detailed studies.

{\bf Conclusion.}--In this letter, we considered the scenario that the left-handed third generation quark doublet $q_L = (t_L, b_L)^T$  are massless bound state of the strong dynamics, using  the minimal coset $SO(5)/SO(4)$ as an example. We studied the constraints on our model from the EWPT ($S$, $T$-parameters and $\delta g_{Lb}$) and direct searches at the LHC. Instead of  the di-boson final state in the case of narrow spin-1 resonances in the Minimal Composite Higgs Model, the smoking gun signature of our model is the broad resonances in the $t\bar{t}$, $b\bar{b}$, $\ell^+\ell^-$, $t \bar{b}/\bar{t}b$, and four top final channels. We have recast the searches at the present LHC and made projections at the HL-LHC. We find that $t\bar{t}$ is comparable to the di-lepton channel in our model and the SSDL from the four top channel can probe the large $g_\rho$ region. Further studies, taking into account additional information such as top angular distribution and polarization, are needed to fully optimize the search for such broad composite resonances. 

{\bf Acknowledgment.}--We would like to thank Roberto Contino, Jiayin Gu for useful discussions.  LTW is supported by the DOE grant DE-SC0013642. DL is supported in part by the U.S. Department of Energy under Contract No. DE-AC02-06CH11357. KPX is supported in part by the National Research Foundation of Korea under grant 2017R1D1A1B03030820.

\appendix
\section{Analytical formulae for the $S$, $T$-parameters}\label{app}

In this appendix, we list the analytical formulae for the $S$, $T$-parameters in our model. We assume that $\xi$ is small and keep the leading terms in $\xi$ expansion. As discussed in the main text, the total contribution can be divided into three classes: the fermion loop, the $\rho$ resonance loop and the Higgs loop with modified Higgs gauge boson coupling:
 \be
\begin{split}
 S &= S_f+ S_\rho+ S_H, \quad T = T_f+ T_\rho+ T_H.\\
\end{split}
\ee
The result for the $S$ parameter reads:
\bea
\label{Sf_analytical}
S_f &=&- \frac{N_cM_t^2\left(4\ln\left(\frac{y_{1R}^2f^2}{M_t^2} \right) - 15\right)}{18\pi y_{1R}^2f^2}\nonumber\\
&-&\frac{N_c\xi\left(4\ln\left(\frac{M_ty_{1R}^5f^5}{\mu^6}\right) + 17-12\ln \frac{\Lambda^2}{M_t^2} \right)}{36\pi}\nonumber\\
S_\rho &=&\frac{4\pi\xi}{g_\rho^2}(1-4\beta_2)
-\frac{\xi}{6\pi}\Big[1+\frac{41}{16}a_\rho^2\nonumber\\
&+&\frac{3}{4}\left(a_\rho^2+28+24\beta_2\left(a_\rho^2\beta_2-a_\rho^2-2\right)\right)\log\frac{\Lambda}{m_\rho}\nonumber\\
&-&\frac{3}{2}\beta_2(9a_\rho^2-4)+\frac{3}{2}\beta_2^2(9a_\rho^2-8)\Big],\nonumber\\
S_H&=&\frac{\xi}{12\pi}\ln\frac{\Lambda^2}{M_h^2},
\eea
while for the $T$-parameter, it reads:
\bea
\label{Tf_analytical}
T_f &=& -\frac{N_cM_t^4\left(6 \ln\left(\frac{y_{1R}^2f^2}{M_t^2}\right)-11\right)}{24\pi M_W^2s_W^2y_{1R}^2f^2}\nonumber\\
&+&\frac{N_cM_t^2\xi\left(\frac{3}{2}\ln\left(\frac{y_{1R}^2f^2}{M_t^2}\right)-5\right)}{24\pi M_W^2s_W^2} -\frac{N_cy_{1R}^2v^2\xi}{96\pi M_W^2s_W^2}.\nonumber\\
T_\rho &=& \frac{9\xi}{32\pi c_W^2}a_\rho^2\left[\left(1-\frac{8}{3}\beta_2^2\right)\log\frac{\Lambda}{m_\rho}+\frac{3}{4}-\frac{4}{3}\beta_2+\frac{2}{9}\beta_2^2\right], \nonumber\\
T_H &=&-\frac{3}{16\pi}\frac{\xi}{c_W^2}\ln\frac{\Lambda^2}{M_h^2}.
\eea
The cutoff $\Lambda$ is chosen as $4\pi f$. From the formulae, we can see that the IR contribution to the $S$, $T$ from modified Higgs gauge boson coupling are anti-correlated. Since the measurement of the $S$, $T$-parameters is strongly correlated $92\%$, this will put a strong bound on the $\xi \lesssim 0.012$, if there are no other contributions.  In our model, both $S_f$ and $T_f$ tend to be negative and the absolute value of $T_f$ is preferred to be larger than $S_f$. Since $\rho$ contribution can be positive, adding the $\rho$ contribution can relax the bound a little bit.

\bibliographystyle{apsrev}
\bibliography{references}

\end{document}